%% file: 0.head.tex
\newcommand{\ie}{\emph{i.e., }}
\newcommand{\eg}{\emph{e.g., }}

\newcommand{\wrt}{\emph{w.r.t. }}

\documentclass[journal]{IEEEtran}
\usepackage{url}
\usepackage{graphicx}
\usepackage{amsmath}
\usepackage{amssymb}
\usepackage{mathrsfs}
\usepackage{multirow}
\usepackage{bm}
\usepackage[justification=centering]{caption}

\ifCLASSOPTIONcompsoc
\usepackage[caption=false, font=normalsize, labelfont=sf, textfont=sf]{subfig}
\else
\usepackage[caption=false, font=footnotesize]{subfig}
\fi

\ifCLASSINFOpdf
\else
\fi
\begin{document}
\title{Hierarchical Attention Network for Visually-aware Food Recommendation}

\author{Xiaoyan Gao,
        Fuli Feng,
        Xiangnan He,
        Heyan Huang,\\
        Xinyu Guan,
        Chong Feng,
        Zhaoyan Ming,
        and~Tat-Seng~Chua
\thanks{The work was supported in part by the National Natural Science Foundation of China (Key Program) under Grant No.61751201 and NExT++ project supported by the National Research Foundation, Prime Ministers Ofﬁce, Singapore under its IRC@Singapore Funding Initiative.}
\thanks{X. Gao, H. Huang and C. Feng are with Beijing Engineering Research Center of High Volume Language Information Processing and Cloud Computing Applications, School of Computer, Beijing Institute of Technology, Beijing, China, 100081. E-Mail: xygao@bit.edu.cn, hhy63@bit.edu.cn, and fengchong@bit.edu.cn}
\thanks{F. Fu, Z. Ming and T.-S. Chua are with School of Computing, National University of Singapore, Singapore, 117417. E-Mail: fulifeng93@gmail, dcsming@nus.edu.sg, and dcscts@nus.edu.sg}
\thanks{X. He is with the University of Science and Technology of China, Hefei, Anhui, China, 230031. E-mail: xiangnanhe@gmail.com}
\thanks{X. Guan is with Systems Engineering Institute, Xi'an Jiaotong University, Xi'an, China, 710049. E-Mail: xinyu\_guan@foxmail.com}
\thanks{Heyan Huang is the corresponding author.}
\thanks{This work is done during Xiaoyan Gao's internship in National University of Singapore (NUS), supervised by Fuli Feng and Xiangnan He.}}

\markboth{IEEE TRANSACTIONS ON MULTIMEDIA, ~VOL.~X, NO.~XX, MONTH~YEAR}%
{Shell \MakeLowercase{\textit{et al.}}: Bare Demo of IEEEtran.cls for IEEE Journals}

\maketitle
\begin{abstract}
Food recommender systems play an important role in assisting users to identify the desired food to eat. Deciding what food to eat is a complex and multi-faceted process, which is influenced by many factors such as the ingredients, appearance of the recipe, the user's personal preference on food, and various contexts like what had been eaten in the past meals. In this work, we formulate the food recommendation problem as predicting user preference on recipes based on three key factors that determine a user's choice on food, namely, 1) the user's (and other users') history; 2) the ingredients of a recipe; and 3) the descriptive image of a recipe. To address this challenging problem, we develop a dedicated neural network-based solution \emph{Hierarchical Attention based Food Recommendation} (HAFR) which is capable of: 1) capturing the collaborative filtering effect like what similar users tend to eat; 2) inferring a user's preference at the ingredient level; and 3) learning user preference from the recipe's visual images. To evaluate our proposed method, we construct a large-scale dataset consisting of millions of ratings from AllRecipes.com. Extensive experiments show that our method outperforms several competing recommender solutions like Factorization Machine and Visual Bayesian Personalized Ranking with an average improvement of 12\%, offering promising results in predicting user preference on food. Codes and dataset will be released upon acceptance.
\end{abstract}

\IEEEpeerreviewmaketitle
\input{1.introduction.tex}

\input{2.motivation}
\input{3.method}
\input{4.evaluation}

\input{5.relatedwork}

\section{Conclusion}\label{Sec:Conclu}
In this work, we proposed a Hierarchical Attention based Food Recommendation (HAFR) system to infer users' preference over recipes for food recommendation. The HAFR aims to learn more comprehensive recipe representation via jointly leveraging user-recipe interaction history, food image, and food ingredients with a hierarchical attention. We collected a large-scale dataset for food recommendation and conducted extensive experiments, demonstrating that HAFR consistently outperforms the state-of-the-art models. Besides, the ablation experiments demonstrate the usefulness of aggregating recipe information in a hierarchical fashion.

In future, we are interested in exploring the following directions. 1) We plan to incorporate healthy and nutrition factors into the food recommendation framework, so that the recommender would guide users to a healthier eating style. 
2) We will explore the relations (\eg replaceable and complementary relations) between ingredients to further enhance the recipe representation. 3) We will incorporate side information of users (\eg gender, age, location and culture) into the food recommendation framework to further improve its performance. 4) We will consider that each recipe contains multiple images instead of one image to enrich recipe representation. 5) We are interested in applying the proposed hierarchical attention into recommendation of items that contain image descriptions and a list of components such as restaurants.

\appendices

\ifCLASSOPTIONcaptionsoff
  \newpage
\fi

\bibliography{bibliography} 
\bibliographystyle{IEEEtran}


\begin{IEEEbiography}[{\includegraphics[width=1in,height=1.25in,clip,keepaspectratio]{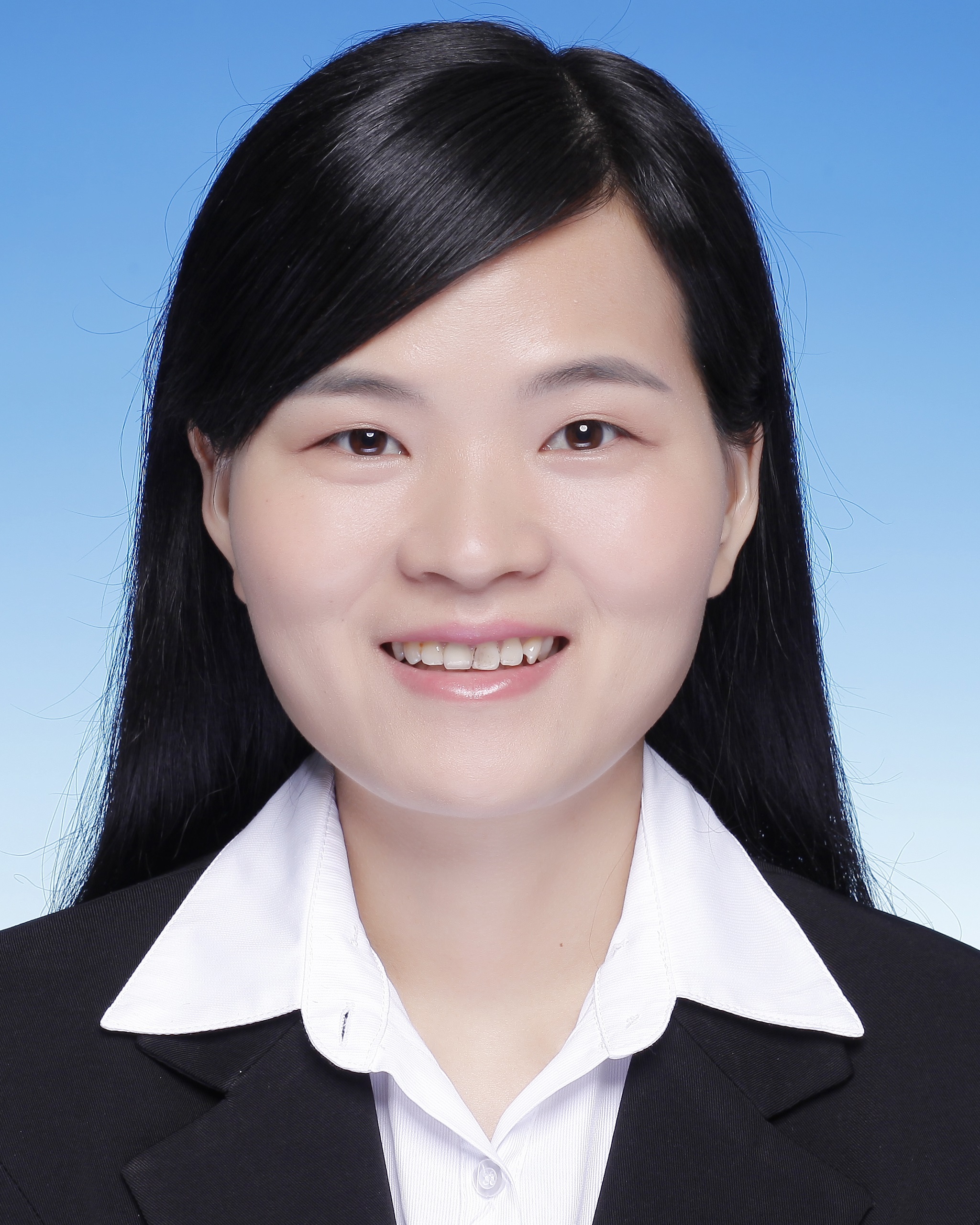}}]{Xiaoyan Gao} is currently pursuing the Ph.D. degree in the School of Computer Science, Beijing Institute of Technology. She received the B.E. degree in software engineering from Fuzhou University, Fujian, in 2012. Her research interests include recommendation, question answering, and natural language processing.
\end{IEEEbiography}

\begin{IEEEbiography}[{\includegraphics[width=1in,height=1.25in,clip,keepaspectratio]{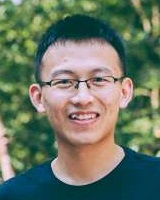}}]{Fuli Feng} is a Ph.D. student in the School of Computing, National University of Singapore. He received the B.E. degree in School of Computer Science and Engineering from Baihang University, Beijing, in 2015. His research interests include information retrieval, data mining, and multi-media processing. He has over 10 publications appeared in several top conferences such as SIGIR, WWW, and MM. His work on Bayesian Personalized Ranking has received the Best Poster Award of WWW 2018. Moreover, he has been served as the PC member and external reviewer for several top conferences including SIGIR, ACL, KDD, IJCAI, AAAI, WSDM etc.
\end{IEEEbiography}

\begin{IEEEbiography}[{\includegraphics[width=1in,height=1.25in,clip,keepaspectratio]{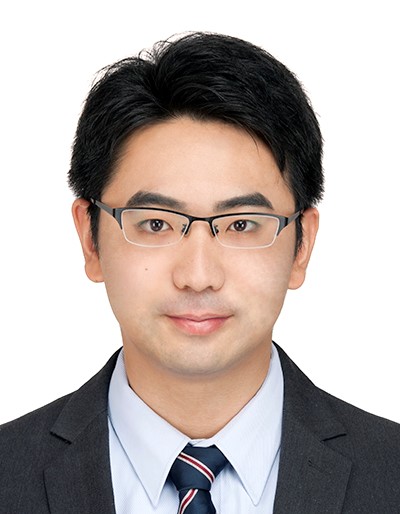}}]{Xiangnan He} is currently a professor with the University of Science and Technology of China (USTC). He received his Ph.D. in Computer Science from National University of Singapore (NUS) in 2016, and did postdoctoral research in NUS until 2018. His research interests span information retrieval, data mining, and multi-media analytics. He has over 50 publications appeared in several top conferences such as SIGIR, WWW, and MM, and journals including TKDE, TOIS, and TMM. His work on recommender systems has received the Best Paper Award Honourable Mention in WWW 2018 and ACM SIGIR 2016. Moreover, he has served as the PC member for several top conferences including SIGIR, WWW, MM, KDD etc., and the regular reviewer for journals including KDE, TOIS, TMM, TNNLS etc.
\end{IEEEbiography}

\begin{IEEEbiography}[{\includegraphics[width=1in,height=1.25in,clip,keepaspectratio]{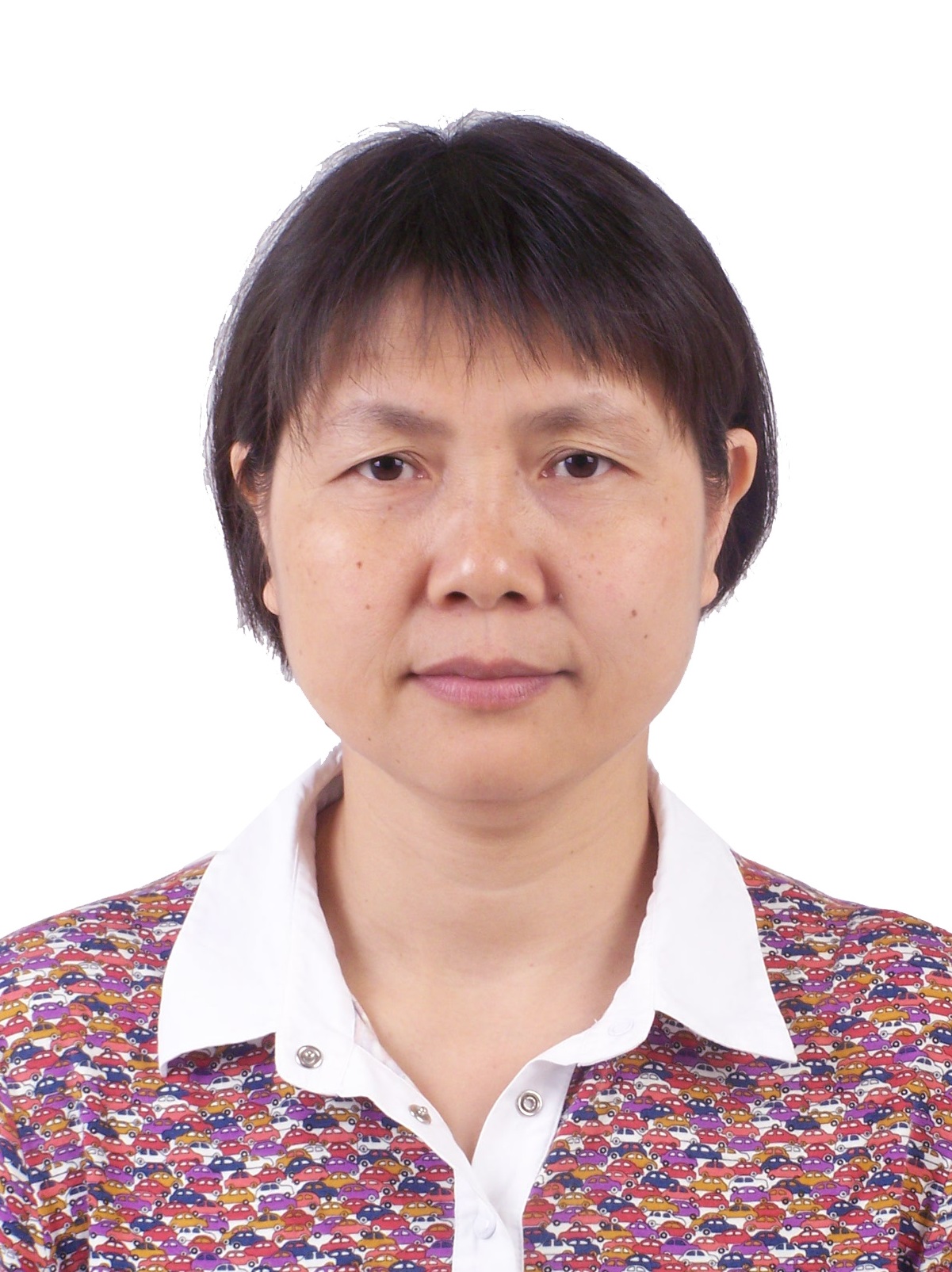}}]{Heyan Huang} received the BE degree in computer science from Wuhan University, in 1983, the ME degree in computer science and technology from the National University of Defense Technology, in 1986, and the PhD degree from the China Academy of Sciences, Institute of Computer Technology, in 1989. Now, she is a professor, doctoral tutor, and president in School of Computer, Beijing Institute of Technology, and the director of the Research Center of High Volume Language Information Processing and Cloud Computing. Her current research interest mainly focus on natural language processing.
\end{IEEEbiography}

\begin{IEEEbiography}[{\includegraphics[width=1in,height=1.25in,clip,keepaspectratio]{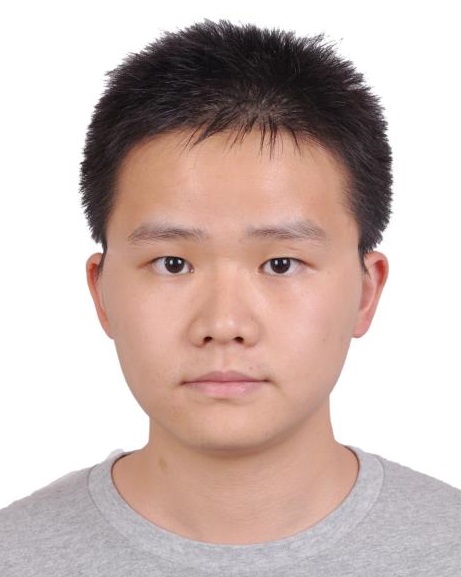}}]{Xinyu Guan} received the B.Sc. degree in automation science and technology from Xi'an Jiaotong University, China, in 2013, where he is currently pursuing the Ph.D. degree with the Systems Engineering Institute. His major research interests are recommender systems and natural language processing.
\end{IEEEbiography}

\newpage
\begin{IEEEbiography}[{\includegraphics[width=1in,height=1.25in,clip,keepaspectratio]{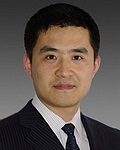}}]{Chong Feng} received the PhD degree from the University of Science and Technology of China, in 2005. Now, he is an associate professor in the School of Computer in Beijing Institute of Technology. His current research interests mainly focus on information extraction and sentiment analysis.
\end{IEEEbiography}

\begin{IEEEbiography}[{\includegraphics[width=1in,height=1.25in,clip,keepaspectratio]{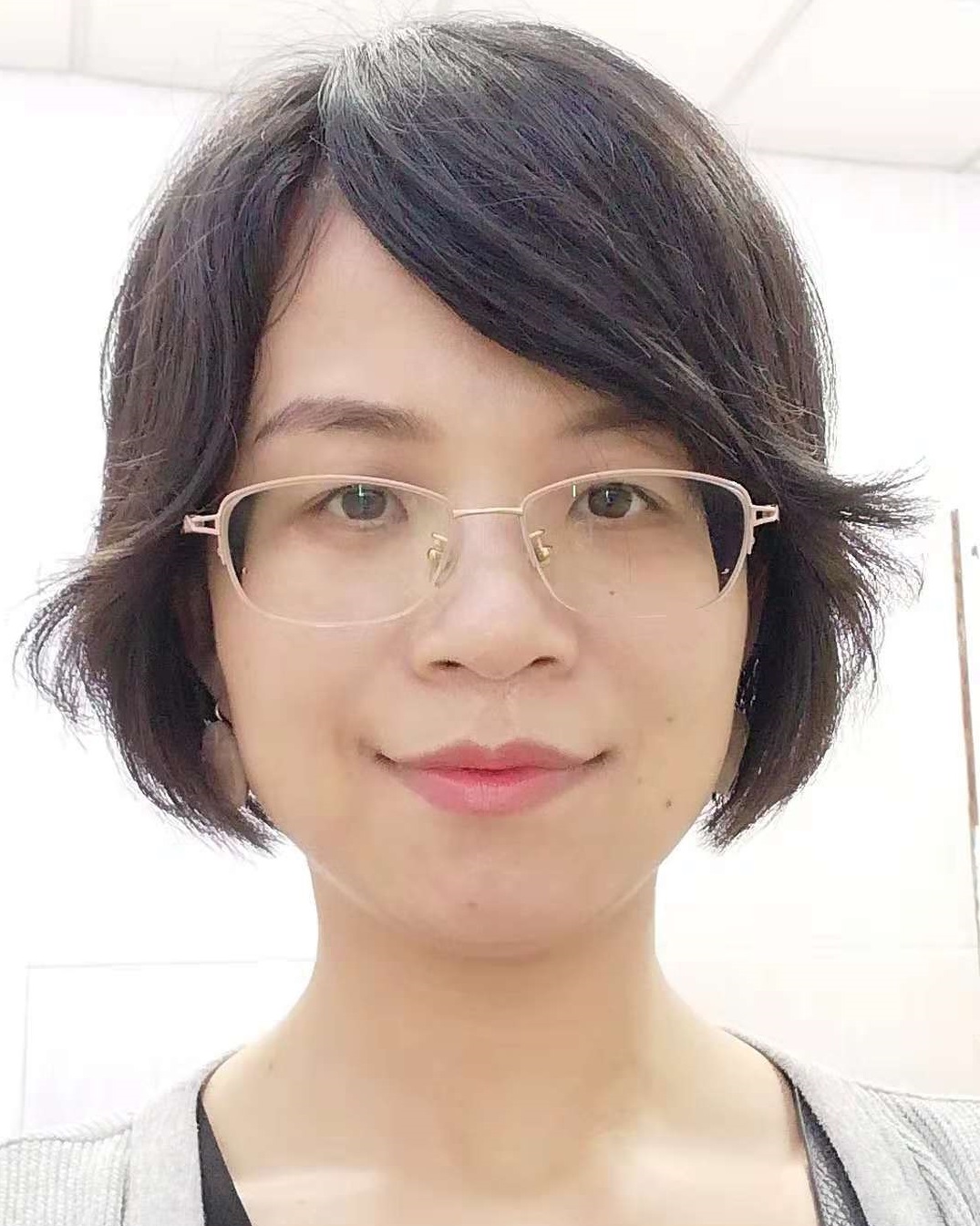}}]{Zhaoyan Ming} is a senior research scientist at School of Computing, National University of Singapore. She is leading the research and development and the collaboration with the medical partners of the wellness project. Her expertise lies in the intersection of knowledge graph and artificial intelligence. She has a strong interest and passion in applying her expertise in machine intelligence to improving people's health. 
\end{IEEEbiography}

\begin{IEEEbiography}[{\includegraphics[width=1in,height=1.25in,clip,keepaspectratio]{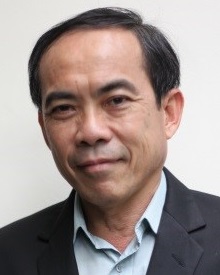}}]{Tat-Seng Chua} is the KITHCT Chair Professor
at the School of Computing, National University
of Singapore. He was the Acting and Founding
Dean of the School during 1998-2000. Dr
Chuas main research interest is in multimedia
information retrieval and social media analytics.
In particular, his research focuses on the extraction,
retrieval and question-answering (QA)
of text and rich media arising from the Web and
multiple social networks. He is the co-Director of
NExT, a joint Center between NUS and Tsinghua
University to develop technologies for live social media search. Dr Chua is the 2015 winner of the prestigious ACM SIGMM award for Outstanding Technical Contributions to Multimedia Computing, Communications and Applications. He is the Chair of steering committee of ACM International Conference on Multimedia Retrieval (ICMR) and Multimedia Modeling (MMM) conference series. Dr Chua is also the General Co-Chair of ACM Multimedia 2005, ACM CIVR (now ACM ICMR) 2005, ACM SIGIR 2008, and ACM Web Science 2015. He serves in the editorial boards of four international journals. Dr. Chua is the co-Founder of two technology startup companies in Singapore. He holds a PhD from the University of Leeds, UK.
\end{IEEEbiography}




\end{document}

%% file: 1.introduction.tex
\section{Introduction}\label{Sec:Intro}
\IEEEPARstart{T}{he} advent of the Internet and mobile technologies facilitates people to access information at any time and place. People's lifestyles have been changed profoundly with the spread of online services, such as social media, E-commerce and various lifestyle Apps. For example, when friends have dinner together, they may take pictures of food they like and share the pictures in social media; when deciding what to eat, users may resort to food-related Apps for recommendations. 

\begin{figure}[!t]
	\centering
	\includegraphics[width=0.475\textwidth]{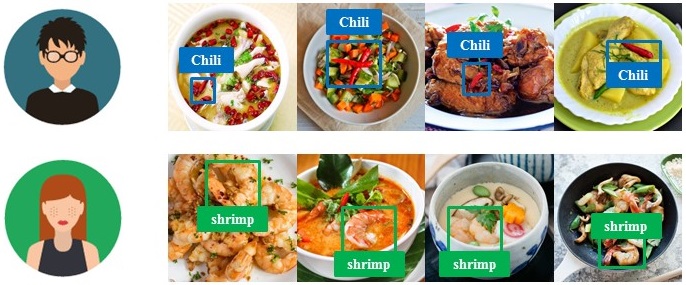}	
	\caption{An example of personal tastes of users on ingredients, in which the man (first row) and the woman (second row) enjoy chili and shrimp, respectively.}
	\label{fig:intro-example}
	\vspace{-0.5cm}
\end{figure}

``Food is the first necessity of the people''. Because of the importance of food in everyday life, food recommender systems have become an indispensable component in many lifestyle services, and are often touted as potential means to affect people towards a healthy lifestyle~\cite{trattner2017investigating}. To build a successful food recommender system, the first step is to accurately understand a user's food preference~\cite{harvey2012learning}. Even for building health-driven food services~\cite{ge2015health}, only when the recommended food (or recipes\footnote{We interchangeably use the term ``food'' and ``recipe'' which refer the same thing in this paper.}) meet the taste of a user, the user can be persuaded to follow the recommendation.  

In this work, we focus on the fundamental problem in building food recommender systems --- inferring user preference on food. While many efforts have been devoted to recommendation techniques~\cite{he2017neural,APR}, they largely focused on the general domains of Movies or E-Commerce products, and relatively little attention has been paid to the specific domain of food. We argue that food is a special class of items that requires dedicated methods to predict user preference on food:
\begin{itemize}
    \item Food is not atomic --- a recipe usually consists of multiple ingredients. In most cases, people prefer a recipe or dish, just because it contains the ingredients they like to eat. An example is shown in Figure \ref{fig:intro-example}, where we sample two users and show their recent four food choices. We find that the first user (male, first row) likes spicy food, as evidenced by the frequently appearance of chili in his chosen food. In contrast, the second user (female, second row) likes seafood, as evidenced by the shrimp in her food choices. As such, to accurately learn user preference on food, it is insufficient to represent a recipe as an ID and simply run collaborative filtering (CF) algorithms like matrix factorization~\cite{FastMF}.
    \item Food image conveys rich information beyond ingredients, being crucial in affecting a user's preference on the food. For example, different cutting and cooking methods can make the same ingredient taste quite differently. ``A picture is worth a thousand words'' --- it is more intuitive (and often easier) for a user to determine the taste of a recipe from its image than from its textual description. As such, it is crucial to properly account for the visual semantics in food image to provide quality recommendation service. 
\end{itemize}

To the best of our knowledge, none of the existing work has approached food recommendation with a comprehensive consideration of all the above-mentioned factors. Existing work has either adopted collaborative filtering which is limited to user-food interaction modeling~\cite{harvey2012learning}, or performed content-based filtering based on the ingredients~\cite{freyne2010intelligent} or image features~\cite{yang2015plateclick}. 
We argue that both user-food interactions and the content features (\ie ingredients and food image) are highly important in food recommendation, such that they should be carefully coupled to infer users' preference on food. Specifically, different users may consume the same food because of different ingredients, and the impact of content features and collaborative effects could also vary for different users. These properties need to be explicitly captured and jointly modeled in the method. 

Targeting at developing dedicated recommendation solution for the food domain, we make the following contributions in this paper:
\begin{itemize}
\item First, we present a new problem formulation, aiming to comprehensively incorporate the key factors that affect a user's food decision-making process, including user-food interaction history, food ingredients, and food image. 
\item Second, we propose a new neural network solution named HAFR, leveraging the strong representation capability of multi-layer neural network and the interpretability of attention mechanism to jointly model the three key factors. 
\item Lastly, we contribute a new dataset for evaluating this task, which is constructed from AllRecipes.com containing over a million ratings. We evaluate our HAFR method on this dataset and demonstrate superior performance over competing recommendation methods\footnote{The dataset is released at: https://www.kaggle.com/elisaxxygao/foodrecsysv1}. 
\end{itemize}

In the remainder of this paper, we first present the problem formulation in Section \ref{sec:pre}, followed by elaborating the method in Section \ref{sec:method} and experimental results in Section \ref{sec:experiments}. Lastly, we discuss related work in Section \ref{sec:related}, conclude the paper, and envision some future work in Section \ref{Sec:Conclu}. 


%% file: 2.motivation.tex
\section{Problem Formulation}\label{sec:pre}
The problem setting of food recommendation is to predict a user's preference on recipes from: 1) \textit{historical user-recipe interactions}; 2) \textit{recipe image}; and 3) \textit{recipe ingredients}. Given a set of users ($\mathcal{U}$) and recipes ($\mathcal{I}$), we use a binary matrix $\bm{Y} \in \mathbb{R}^{M \times N}$ to denote the user-recipe interactions, where $M$ and $N$ denote the number of users and recipes, respectively. Each entry $y_{ui}$ denotes whether a user $u$ has interacted with (\ie rated) a recipe $i$, which is defined as:
\begin{equation}\label{eq:Y_uir}
y_{ui} =\left\{
\begin{aligned}
	1 & , \quad\textrm{if user $u$ consumed recipe $i$;} \\
	0 & , \quad\textrm{otherwise.}
\end{aligned}
\right.
\end{equation}
Besides the ID $i$, a recipe also contains an image $\bm{V}_i$\footnote{This work considers that each recipe contains one image only and leaves the extension to multiple images as future work.} and a set of ingredients $\bm{g}_i$. $\bm{V}_i$ denotes the raw image feature, $\bm{g}_i \in \mathbb{R}^{K}$ is a multi-hot encoding with $g_{i}^{t} = 1$ denoting that ingredient $t$ is in recipe $i$, where $K$ is the number of ingredients occurred in $\mathcal{I}$, and $t$ denotes the ingredient ID. Then, our task is to learn an interaction function $\hat{y}_{ui} = f(u, i, \bm{g}_i, \bm{V}_i)$ which estimates the probability that user $u$ would interact with recipe $i$, which is formally defined as:

\textbf{Input:} Users $\mathcal{U}$, recipes $\mathcal{I}$, user-recipe interactions $\bm{Y}$, recipe ingredients $[\bm{g}_1, \cdots, \bm{g}_N]$, and recipe images $[\bm{V}_1, \cdots, \bm{V}_N]$. 

\textbf{Output:} An interaction function $\hat{y}_{ui}=f(u, i, \bm{g}_i, \bm{V}_i)$, which outputs the probability that user $u$ would consume recipe $i$.

%% file: 3.method.tex
\section{Methodology}\label{sec:method}
%
\begin{figure}[t]
	\includegraphics[width=0.48\textwidth]{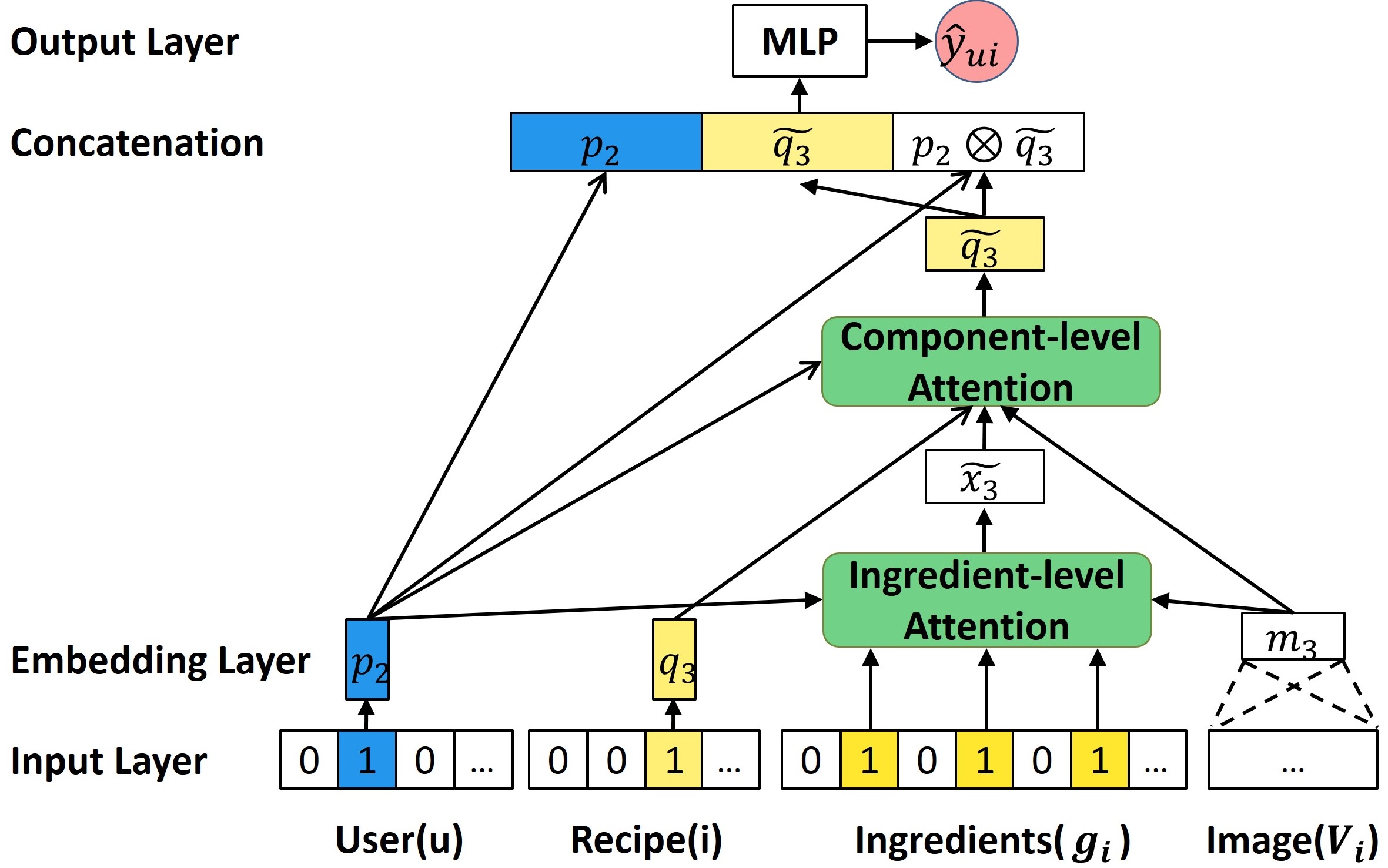}
	\caption{Illustration of the proposed HAFR.}
	\label{fig:framework}
	\vspace{-0.5cm}
\end{figure}
In this section, we introduce the proposed HAFR for solving the food recommendation problem. Different from existing food recommender systems that either mine the user-recipe interactions or analyze recipe contents \cite{trattner2017investigating,yang2015plateclick}, we explicitly incorporate ingredients and visual image as key 
information to enhance the representation of recipe and better infer users' preference over recipes.
We present the architecture of HAFR model in Figure \ref{fig:framework}, which consists of three main components:
\begin{itemize}
    \item \textbf{Embedding layer} encodes user-recipe interaction (user ID and recipe ID), recipe image, and recipe ingredients as the embedding of user-recipe, image, and ingredients, respectively.
    \item \textbf{Hierarchical attention} dynamically aggregates the embeddings into a more comprehensive recipe representation ($\widetilde{\bm{q}_3}$) by accounting for the preference of the target user.
    \item \textbf{Output layer} composes the vector representation of the user and the recipe, which is fed into a multi-layer perceptron (MLP) to estimate $\hat{y}_{ui}$.
\end{itemize}

The remainder of this section is organized to elaborate each of the three components. 
%
%
\subsection{Embedding Layer}
\subsubsection{User-recipe interaction}
Embedding-based models \cite{he2017neural,wang2018tem} associate each user and item with an embedding (i.e., a real-valued vector), which has become the mainstream to model user-item interactions in recommendation \cite{chen2017attentive}. 
Considering their recent success, we also project the user ID and recipe ID (\ie $u$ and $i$) into an embedding, which is expected to encode the collaborative signals among users and recipes. 
Formally, we first encode user ID $u$ (item ID $i$) into an one-hot encoding $\bm{e}_u \in \mathbb{R}^M$ ($\bm{e}_i \in \mathbb{R}^N$) with zero values except the $u$-th ($i$-th) entry.
We then employ two embedding layers to project the one-hot encodings into embeddings of user and item, which are formulated as,
\begin{align}
\bm{p}_u = \bm{P} \bm{e}_u, ~~ \bm{q}_i = \bm{Q} \bm{e}_i,
\end{align}
where $\bm{p}_u$ and $\bm{q}_i \in \mathbb{R}^D$ are the user and recipe embeddings, respectively; $\bm{P} \in \mathbb{R}^{D \times M}$ and $\bm{Q} \in \mathbb{R}^{D \times N}$ are all embedding parameters to be learned, where $D$ is the embedding dimension. Note that we can also view an embedding layer as a look-up operation, which retrieves an embedding from the parameter matrix of the embedding layer, considering that $\bm{p}_u$ is the $u$-th column of $\bm{P}$. To simplify the presentation, in the following sections, we denote the operation of one-hot encoding and embedding layer as a look-up function $\bm{p}_u = lu(u, \bm{P})$.

\subsubsection{Recipe Image}
As a significant information carrier, recipe image presents rich information, playing an important role in affecting users' preference and selection on recipes~\cite{elsweiler2017exploiting,chokr2017calories}.
To enhance the representation of recipe and facilitate the inference of user preference, we incorporate the visual signals from recipe image. 
Inspired by its success in many computer vision tasks, we employ the ResNet-50~\cite{he2016deep} to extract visual features $\bm{v}_i$ from the raw image $\bm{V}_i$. 
Following~\cite{chen2018deep}, we use the output of the \emph{pool5} layer as the visual features, which is a 2,048 dimension vector. 
Note that the ResNet-50 is pre-trained on ImageNet \cite{deng2009imagenet} and fine-tuned via classifying raw recipe images into their associated categories (\eg chicken rice).

Considering that the embedding size of user and item is typically much smaller than 2,048, we project $\bm{v}_i$ into a latent space of dimension $D$ through a mapping layer, of which the formulation is,
\begin{equation}
    \bm{m}_i = \bm{W} \bm{v}_i + \bm{b},
\end{equation}
where $\bm{m}_i \in \mathbb{R}^{D}$ is termed as image embedding, $\bm{W} \in \mathbb{R}^{D \times 2048} $ and $\bm{b} \in \mathbb{R}^{D}$ are parameters to be learned. Note that we only use one mapping layer for the projection instead of stacking multiple layers for the consideration of model complexity.

\subsubsection{Recipe Ingredients}
\begin{figure}[t]
	\centering
	\includegraphics[width=0.32\textwidth]{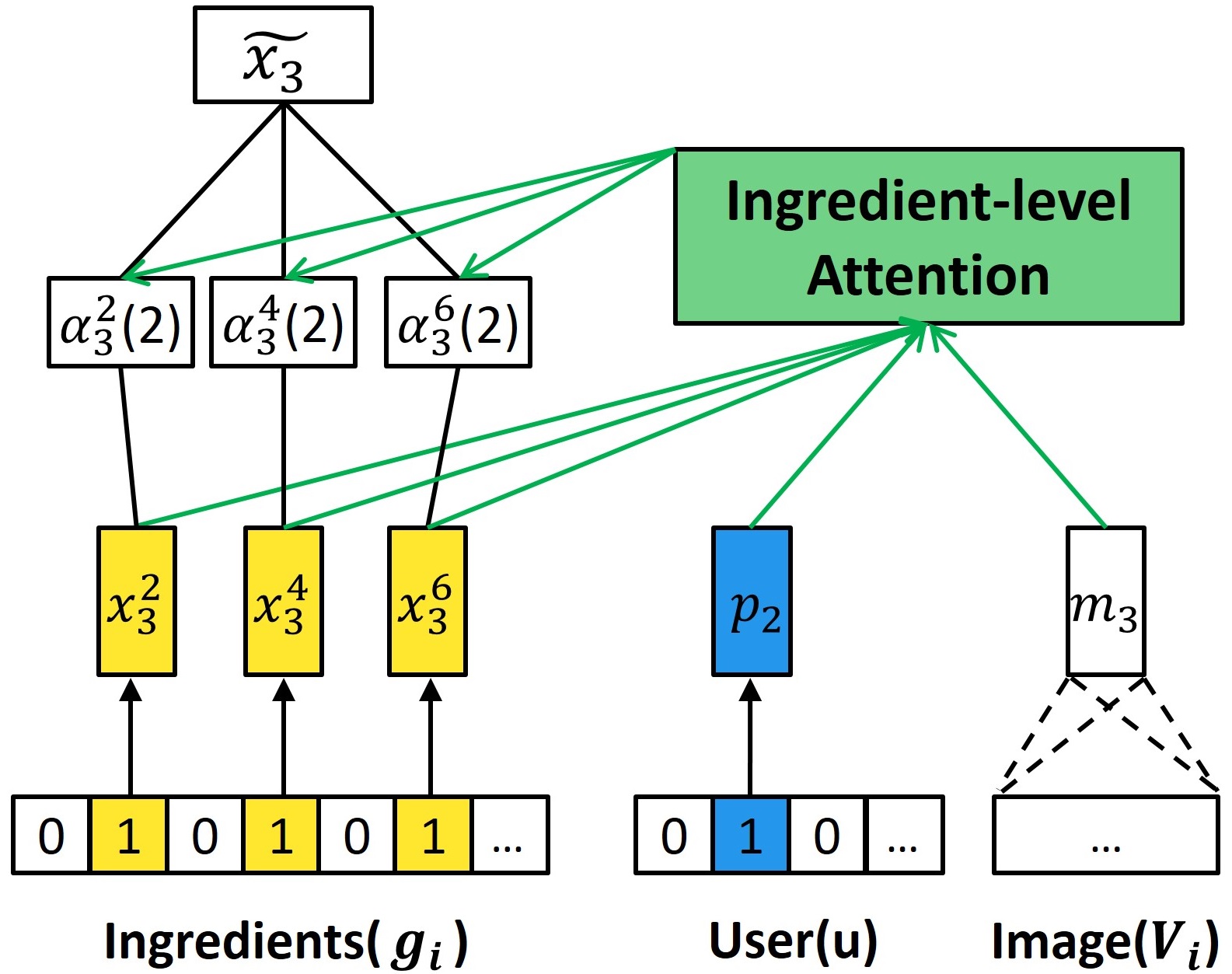}
	\caption{Details of Ingredient-level Attention module.}
	\label{fig:attention}
	\vspace{-0.5cm}
\end{figure}
Previous work adopting content-based filtering for food recommendation has shown that ingredients of a recipe could be important clues for inferring user preference over food~\cite{freyne2010intelligent}. For instance, a user would consume a food item simply because it contains his/her favourite ingredients. As such, we further encode the ingredients of a recipe to enrich its representation. Note that the multi-hot encoding $\bm{g}_i \in \mathbb{R}^{K}$ is a sparse ($K$ is typically larger than ten thousand) vector with binary values. To overcome the sparsity problem, we also learn an embedding for each ingredient. Formally, we denote the ingredient embeddings as $\bm{X} = [\bm{x}^{1}, \cdots, \bm{x}^{K}] \in \mathbb{R}^{D \times K}$, where $\bm{x}^{k}$ is the embedding of ingredient $k$. From $\bm{X}$, we then obtain the embeddings of ingredients occurred in recipe $i$, \ie
\begin{align}
   \mathcal{X}_i = \{\bm{x}^{k} | g_i^k = 1\},\text{ where } \bm{x}^{k} = lu(k, \bm{X}),
\end{align}
recall that $\bm{g}_i$ is the multi-hot encoding of the ingredient list.

\subsection{Hierarchical Attention}
The information of a recipe inherently has a hierarchical structure with two levels: \textit{component-level} and \textit{ingredient-level}. That is, a recipe consists of three components: interaction signal ($\bm{q}_{i}$), image features ($\bm{m}_{i}$), and an ingredient list ($\mathcal{X}_i$). The ingredient list further contains multiple ingredients. To leverage the hierarchical structure, we devise a hierarchical attention with \textit{component-level attention} and \textit{ingredient-level attention} to fuse the recipe information to obtain a better recipe representation.

\subsubsection{Ingredient-level Attention}
Considering that different recipes would contain different numbers of ingredients, we employ a pooling operation to aggregate the ingredient embeddings into a single vector, so as to mitigate the problem of variant length. Our first intuition is to equally fuse the ingredient embeddings with an average pooling,
\begin{equation}
    \widetilde{\bm{x}_{i}} = \frac{1}{|\mathcal{X}_i|} \sum_{g_i^k = 1}  {\bm{x}_{i}^k},
\end{equation}
where $|\mathcal{X}_i|$ denotes the size of $\mathcal{X}_i$.
However, the average pooling lacks the flexibility to dynamically adjust the weights of ingredients. This flexibility is particularly useful when a user is more interested in certain ingredients. 
For instance, a user enjoys chili might have higher probability to select recipes containing chili. As such, the embedding of chili may contribute more than other ingredients to the compressed embedding of the given user. Therefore, we argue that the embedding of an ingredient should be fused with adaptive weights to capture users' dynamic preference over ingredients. We formulate the adaptive fusion as:
\begin{equation}
\widetilde{\bm{x}_{i}} = \sum_{k}  {\alpha^k_i (u)\bm{x}_{i}^k},\text{ where } g_i^k = 1.
\end{equation}

Inspired by the recent success of attention mechanism~\cite{bahdanau2015neural,xiao2017attentional}, 
which allows different parts to contribute differently when compressing them to a single representation, 
%
we propose an \textit{ingredient-level attention} to model the adaptive weight $\alpha^k_i (u)$. 
Specifically, as illustrated in Figure~\ref{fig:attention}, we use a two-layer network to compute $\alpha^k_i (u)$ with user embedding $\bm{p}_u$, image embedding $\bm{m}_i$, and the embedding of the target ingredient $\bm{x}_{i}^{k}$ as input,

\begin{align}
    &\alpha_i^k(u) = \frac{\exp(a_i^k(u))} {\sum_{k}{\exp(a_i^k(u))}} ,\text{ where } g_i^k = 1\\
    &a_i^k(u) = \bm{h}^T tanh(\bm{W}_{1p}\bm{p}_u + \bm{W}_{1m} \bm{m}_i + \bm{W}_{1x} \bm{x}_{i}^k + \bm{b}_{1}),\notag
\end{align}
where $\bm{W}_{1*} \in \mathbb{R}^{D_1 \times D}$, $\bm{b}_{1} \in \mathbb{R}^{D}$, and $\bm{h} \in \mathbb{R}^{D_1}$ are the parameters to be learned, $D_1$ is the size of hidden layer. $tanh$ is a hyperbolic tangent function.
\subsubsection{Component-level Attention}
%
By far, we have three representations ($\bm{q}_i$, $\bm{m}_i$, $\widetilde{\bm{x}_{i}}$) with the same dimension, corresponding to different components of the recipe (\ie ID, recipe image, and recipe ingredients). One straightforward way to jointly account for the three components is late fusion, \ie separately estimating three probabilities from ($\bm{p}_u$, $\bm{q}_i$), ($\bm{p}_u$, $\bm{m}_i$), and ($\bm{p}_u$, $\widetilde{\bm{x}_{i}}$), and fusing them into an overall probability of the interaction between $u$ and $i$. However, this would present issues due to the increase in model complexity since different probabilities might rely on different parameters and network structures so as to be accurately estimated. To simplify the problem, we first aggregate $\bm{q}_i$, $\bm{m}_i$, and $\widetilde{\bm{x}_{i}}$ into a more comprehensive food representation which comprehensively encodes both the user-recipe interactions and content features of the recipe.

Similar to the fusing of ingredient embeddings into $\widetilde{\bm{x}_{i}}$, we aggregate the three components (\ie $\bm{q}_i$, $\bm{m}_i$, and $\widetilde{\bm{x}_{i}}$) via a user-aware adaptive strategy,
\begin{equation}
\widetilde{\bm{q}_{i}} = \sum_{\bm{c}_{\rho} \in \{\bm{q}_i, \bm{m}_i, \widetilde{\bm{x}_{i}}\}} {\beta_{\rho}(u) \bm{c}_{\rho}},
\end{equation}
where $\beta_{\rho}(u)$ is the user-aware coefficient to adjust the importance of different components. Note that we use $\bm{c}_{\rho}$ to represent the component representation to simplify the illustration. The motivation of incorporating the user-aware coefficient is to capture users' dynamic preference over different components. For instance, one user might choose a food for its attractive appearance, whereas another user would prefer food similar to her preferred tastes.  
%
Similar to \textit{ingredient-level attention}, we devise a \textit{component-level attention} module to compute the user-aware coefficients. 
Specifically, the component-level attention module is also a two-layer network formulated as,
\begin{align}
&\beta_{\rho}(u) = \frac{\exp(b_{\rho}(u))}{\sum_{\bm{c}_{n} \in \{\bm{q}_i, \bm{m}_i, \widetilde{\bm{x}_{i}}\}} {\exp(b_n(u))}}, \notag\\
&b_{\rho}(u) = \bm{v}^T tanh(\bm{W}_{2p}\bm{p}_u + \bm{W}_{2\rho} \bm{c}_{\rho} + \bm{b}_{2}),
\end{align}
where the matrices $\bm{W}_{2*}$ and bias $\bm{b}_{2}$ are parameters to be learned.

It's worth mentioning that our design of hierarchical representation learning is inspired from the hierarchical attention network (HAN)~\cite{yang2016hierarchical}. However, our proposed hierarchical attention is different from HAN, which is originally proposed to learn document representations, where a word-level attention extracts important words for sentence representation and a sentence-level attention rewards sentences providing clues for downstream applications \cite{yang2016hierarchical}. Instead of merely calculating the coefficients from the contents (\eg words) like HAN, the ingredient-level and component-level attentions calculate coefficients conditioned on user embedding to capture users' dynamic preference. Moreover, our hierarchical attention aggregates heterogeneous inputs (\ie user-recipe interactions, image, and ingredients) which differs significantly from the homogeneous inputs (\ie words) of HAN.

Furthermore, the proposed hierarchical attention is distinct from HAN in Attentive Collaborative Filtering (ACF)~\cite{chen2017attentive}, which is the first recommendar system leveraging the idea of HAN. Specifically, ACF uses HAN to learn user representation from the images the user has interacted with (the hierarchy is: a user$->$historical images; an image$->$regions). In contrast, we apply HAN to learn the representation of recipe (\ie item) of which the information has a totally different hierarchy. Moreover, we focus on the specific domain of food, while ACF works on the general domain of multimedia content.

\subsection{Output Layer and Training}
\subsubsection{Output Layer}
Until now, given a user-recipe pair ($u, i$) with the associated image and ingredients, we obtain the embeddings for user $\bm{p}_u$ and recipe $\widetilde{\bm{q}_{i}}$. Following \cite{he2017neural}, we compose a representation for the user-recipe pair by concatenating user embedding, recipe embedding, and their element-wise product (${\bm{p}_{u}} \odot \widetilde{\bm{q}_{i}}$). 
We then feed the unified representation into a MLP in order to employ the nonlinear function for modeling the complicated interactions~\cite{he2017neural,wang2017item}. Formally, the output layer estimates the interaction probability between user and recipe as:
\begin{equation}
\hat y_{ui} = \bm{z}^T f(\bm{W}_3 
\begin{bmatrix}
	{\bm{p}_u}     \\
	\widetilde{\bm{q}_i}       \\
	{\bm{p}_u} \odot \widetilde{{\bm{q}_i}} \\
\end{bmatrix} + \bm{b}_3)
\end{equation}

\noindent where, $\bm{W}_3$, $\bm{b}_3$, and $\bm{z}$ denote the weight matrix, bias vector, and weights of output layer, respectively. $f(\cdot)$ is set to the ReLU function, which has empirically shown to work well.

\subsubsection{Objective Function}
Since we address the food recommendation task from the ranking perspective, we employ a pairwise learning method to optimize model parameters. The assumption of pairwise learning is that the model could predict a higher score for an observed interaction than its unobserved counterparts. Specifically, an observed user-recipe interaction is assigned to a target value 1, otherwise 0. We adopt the Bayesian Personalized Ranking (BPR) loss, which has been widely used in recommendation~\cite{chen2017attentive,he2018adversarial}: 
\begin{equation}
\mathcal{L} = \sum_{(u,i,k) \in \mathcal{D_S}} {-ln\sigma(\hat y_{ui} - \hat y_{uk}) + \lambda ||\Theta||^2},
\end{equation}
where $\sigma$ is the logistic (sigmoid) function and $\lambda$ is a model specific regularization hyperparameter. $\Theta$ denotes the parameters in our model. $\mathcal{D_S}$ is the training set, which consists of triples in the form $(u,i,k)$, where $u$ denotes the user together with an interacted recipe $i$ and a non-observed recipe $k$:
\begin{equation}
\mathcal{D_S} = \left \{ 
(u,i,k) | y_{ui} = 1 \wedge  y_{uk} = 0 \right \}
\end{equation}

To optimize the above objective function, we adopt Adagrad~\cite{duchi2011adaptive}, a variant of Stochastic Gradient Descent (SGD) that applies an adaptive learning rate for each parameter. It draws a stochastic sample from all training instances, updating the related parameters towards the negative direction of their gradients. As TensorFlow provides the function of automatic differentiation, we omit the derivation of the derivatives of our model. In this paper, we use the mini-batch version of Adagrad to speedup the training process. In each training epoch, we sample the negative instance (\ie sampling $k$ for ($u$, $i$)) on-the-fly, as is done in BPR~\cite{rendle2009bpr}. 

%% file: 4.evaluation.tex
\section{Experiments}\label{sec:experiments}
We conducted extensive experiments on a real-world dataset to answer the following research questions:

\begin{itemize}
	\item \textbf{RQ1:} How does our proposed model perform as compared with the state-of-the-art models that are designed for food recommendation?
	\item \textbf{RQ2:} Is the proposed hierarchical attention helpful for learning the recipe representation and improving the recommendation performance?
	\item \textbf{RQ3:} How do the models perform with respect to the number of ratings per recipe, \ie the popularity of recipe?
	\item \textbf{RQ4:} What are the key hyper-parameters for HAFR and how do they impact HAFR's performance?
\end{itemize}
In what follows, we first describe the experimental settings, followed by answering the above research questions.

\subsection{Experimental Settings}
\subsubsection{Data Collection.} To the best of our knowledge, existing public food datasets, including Food-101~\cite{bossard2014food} and Yummly-28K~\cite{min2017being}, are not suitable for evaluating recommendation since they lack the records of user-food interactions. 
Therefore, we construct a dataset via collecting data from Allrecipes.com. The website is chosen since it is one of the largest food-oriented social networks with 1.5 billion visits per year. In particular, we crawl 52,821 recipes in 27 different categories, which are posted between the year of 2000 and 2018. For each recipe, we crawl its ingredients, image and the corresponding ratings from users. The ratings are transformed into binary implicit feedback as ground truth, indicating whether the users has interacted with the specific recipe. This is because we focus on modeling users' preference over recipes, where the rating activity reflects a strong interest on the recipe (the user has tried the recipe). 

\begin{figure}[t]
	\hspace{-0.1in}
	\centering
	\subfloat[]{\includegraphics[width=0.24\textwidth]{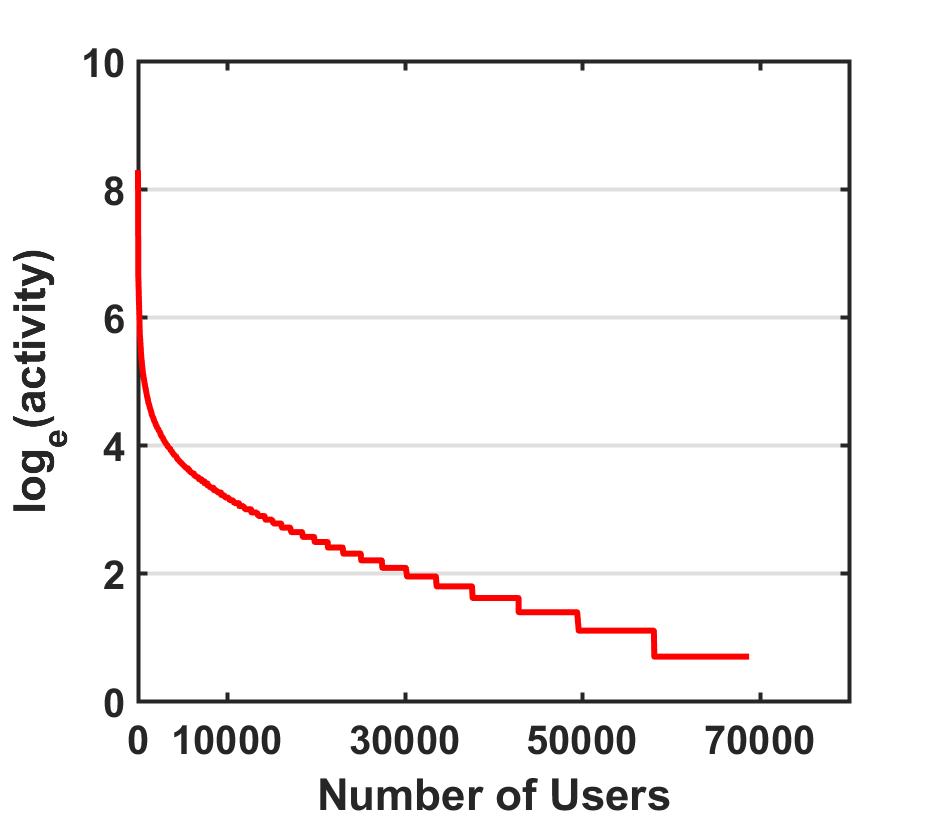}
	\label{fig:POP_user}}
	\hspace{-0.15in}
	\subfloat[]{\includegraphics[width=0.24\textwidth]{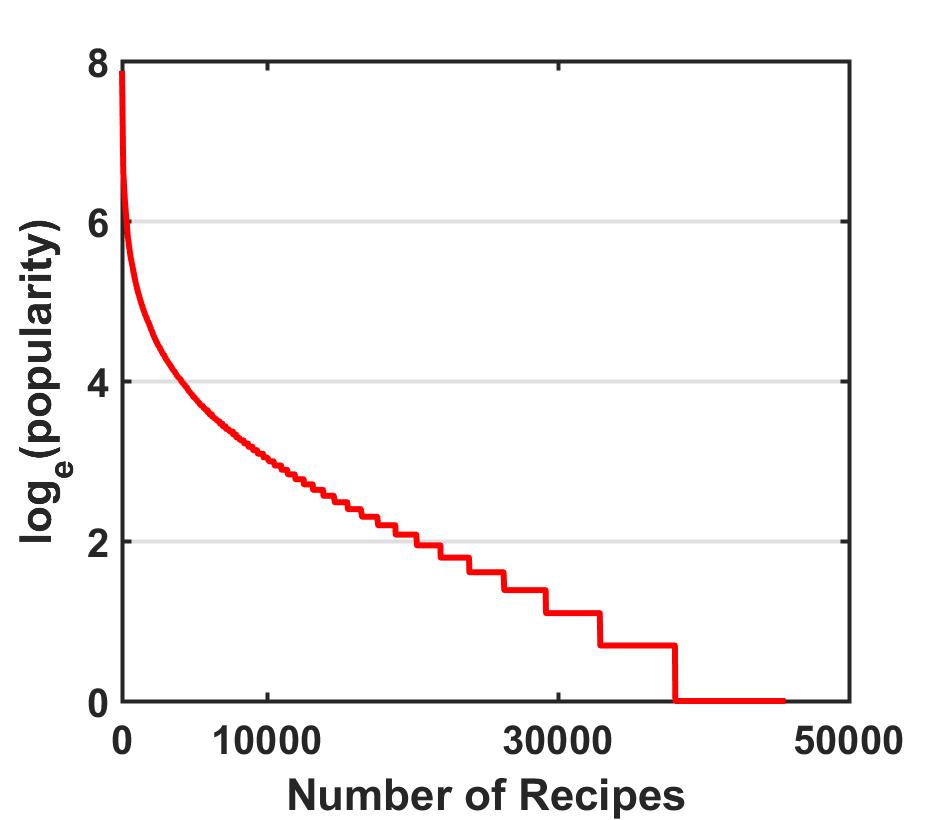}
	\label{fig:POP_item}}

	\caption{The distribution of user activity and recipe popularity over the Allrecipes dataset.}
	\label{Fig:POP}
	\vspace{-0.4cm}
\end{figure}

To evaluate the model performance, we holdout the latest 30\% of interaction history to construct the test set, and split the remaining data into training (60\%) and validation (10\%) sets. We then retain users which occur in both training and test sets, and obtain 68,768 users, 45,630 recipes with 33,147 ingredients and 1,093,845 interactions. Moreover, we depict the distribution of user activity and recipe popularity on the constructed dataset in Figure \ref{Fig:POP}. 
We can see that most recipes only have few ratings, \ie most recipes lie in the long tail of the distribution. Similarly, The majority of users only rates several recipes, which indicates the long tail distribution of user activity. 
Note that the dataset is publicly accessible via: \url{https://www.kaggle.com/elisaxxygao/foodrecsysv1}.


\textbf{Evaluation Protocol.} Given one user and its rated recipes in the test set, we sample 500 negative instances that the user has not interacted. 
Considering the long tail distribution of recipe popularity, we bias the negative sampling towards the popular recipes. Specifically, we calculate the popularity of a recipe $f_i$ on the training set, and sample it with a probability of $f^r_i / \sum_{i' \in \mathcal{I}} f^r_{i'}$. $r$ is a coefficient to control the bias (larger $r$ leads to higher probability for popular recipes), which is empirically set as $r=0.7$. Note that this sampling strategy is unfriendly to methods with popularity bias (\ie potentially predicting popular recipes as positive), and benefit methods that focus on inferring users' preference, which is the key target of food recommendation.

We then estimate the interaction probability of each instance and generate a ranking list. To evaluate the performance, We adopt three popular metrics for recommendation, \emph{AUC}, \emph{NDCG} and \emph{Recall}, to judge the performance of the ranking list:
\begin{itemize}
	\item \textbf{AUC}
	\emph{Area Under the Roc Curve} (AUC) measures the probability that a classifier will rank a randomly chosen positive instance higher than a randomly chosen negative one. 
	\item \textbf{NDCG} \emph{Normalized Discounted Cumulative Gain} (NDCG) assigns higher scores to the hits at higher positions of the ranking list.
	\item \textbf{Recall}
	Recall is the probability that positive items are ranked in top-\emph{k} item recommendation.
\end{itemize}
For all metrics, the values range from 0 to 1 (the higher the better). Note that we report the average score for 10 different repeats of testing, and apply \emph{t}-test to obtain the statistical significance. 

\subsubsection{Baselines.} 
To justify the effectiveness of our proposed model, we compared the performance of our proposed model with the following baselines. Note that all methods are learned by optimizing the same pairwise ranking loss of BPR for fair comparison. The compared recommendation methods are introduced as follows:
\begin{itemize}
    \item \noindent\textbf{LDA~\cite{trattner2017investigating}}. This method is the state-of-the-art food recommendation method. For LDA recommender, users are regarded as documents and recipes as words. Following \cite{trattner2017investigating}, we use the Librec\footnote{https://www.librec.net/} implementation and tune the topic number. 
    
    \item \noindent\textbf{MF-BPR~\cite{rendle2009bpr}}. \emph{Matrix Factorization-Bayesian Personalized Ranking} (MF-BPR) optimizes the standard MF model with the pairwise BPR ranking loss. BPR is a widely used pairwise learning framework for item recommendation. This method is a popular choice for building a CF recommender from implicit feedback. 

    \item 
    \noindent\textbf{FM~\cite{rendle2012factorization}}.
    \emph{Factorization Machine} (FM) estimates the target by modeling all interactions between each pair of features via factorized interaction parameters. In this work, we set the input features as user ID, recipe ID, and ingredients. By using the second-order factorized interactions between features, FM can mimic many specific factorization models such as the standard MF. Therefore, FM has been considered as one of the most effective embedding methods for sparse data prediction.

    \item \noindent\textbf{VBPR~\cite{he2016vbpr}}. The method incorporates visual contents to MF-BPR, using the visual contents as parts of recipe descriptions to predict users' preference over recipes. 
    
    \item \noindent\textbf{FM-VBPR}. To test usefulness of visual features and ingredient features simultaneously, we add visual features into FM framework above, which is projected into an embedding as the user ID, recipe ID, and ingredients. We term this method as FM-VBPR.
\end{itemize} 
Note that the input for MF-BPR and LDA are users and recipes embeddings. Different from the above methods, FM adds ingredient embeddings as input features, VBPR adds visual features as input. Similar with our method HAFR, FM-VBPR considers all features as input, including users, recipes, ingredients and visual features. For hyperparameters tuning, We tune the learning rate and the coefficient for $L_2$ regularization for all models except LDA.
%

\subsubsection{Parameter Settings.} We implement our model and baseline methods in TensorFlow, which would be accessible via \url{https://anonymous.com}. For each method, we select the optimal hyper-parameters \wrt AUC on validation set. For a fair comparison, all models are set with an embedding size of 64 and optimized using the mini-batch Adagrad~\cite{duchi2011adaptive} with a batch size of 512. Besides, we search the learning rate within [0.0001, 0.0005, 0.001, 0.005, 0.01, 0.05] and $\lambda$ for the regularization term among [0.0001, 0.001, 0.01, 0.1, 1]. Note that we use different $\lambda$ for parameters of embeddings ($\bm{P}$, $\bm{Q}$, and $\bm{X}$), mapping layer of image feature, and the MLP of output layer. 
Instead of considering both the two levels of attentions, basic HAFR uses a uniform weight to all entities. 
The proposed HAFR achieves good performance when the learning rate is 0.05 and regularization terms for embedding, image mapping layer, and MLP equal to [0.1, 0.01, 1], respectively. Without special mention, we report the performance of HAFR on this special setting.

\begin{table}[t]
	\begin{center}
	\small
		\caption{Performance of compared methods. $^*$ denotes the statistical significance for $p < 0.05$.}\label{tab:model performance}
		\renewcommand{\arraystretch}{1.1}
		\begin{tabular}{|c|c|c|c|}
			\hline
			\multirow{2}{*}{\bf Methods} & \multicolumn{3}{|c|}{\bf Allrecipes}   \\\cline{2-4}
			&  \bf AUC  & \bf NDCG@10 & \bf Recall@10 \\\hline\hline
			
			MF-BPR    &0.5622  &0.0376 &0.0567  \\\hline
			LDA    &0.5154  &0.0376 &0.0601\\\hline
			FM &0.5710  &0.0396 &0.0607
			\\\hline
			VBPR     &0.5808  &0.0296 &0.0431 \\\hline
			FM-VBPR    &0.5840  &0.0372 &0.0580 \\\hline\hline
			HAFR-non-v &0.6004   &0.0332 &0.0517 \\\hline
			HAFR-non-i &0.6133   &0.0418 &0.0608 \\\hline
			
			HAFR &$\bf{0.6435}^*$ &$\bf{0.0455}^*$ &$\bf{0.0674}^*$\\\hline
		\end{tabular}
		\vspace{-0.5cm}
	\end{center}
\end{table}

\begin{figure}[t]
	\hspace{-0.1in}
	\centering
	{
		\includegraphics[width=0.24\textwidth]{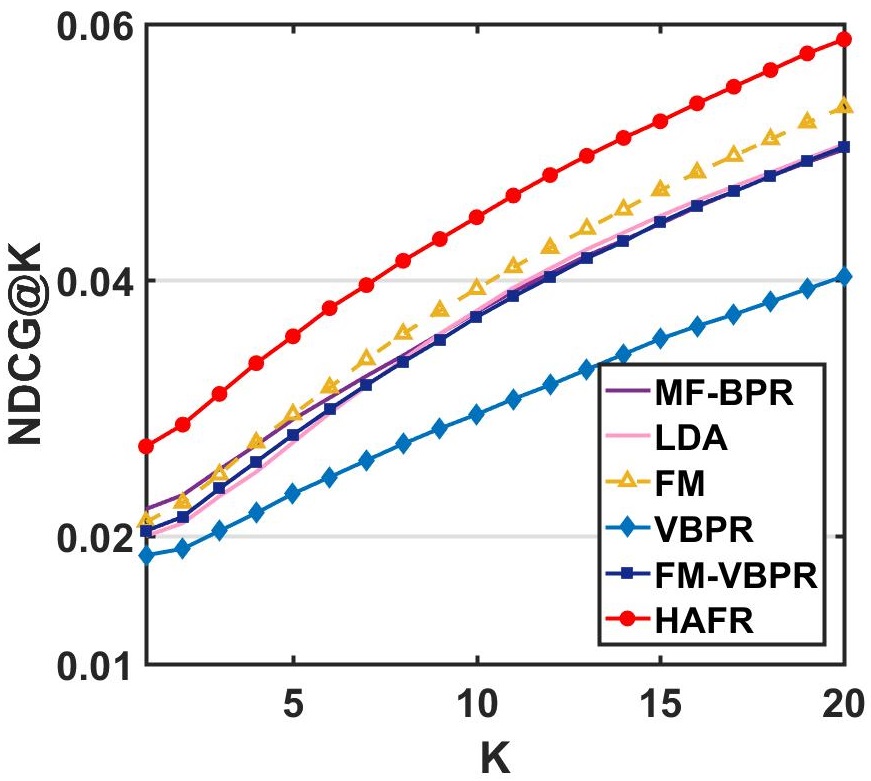}
		\label{fig:RQ_NDCG}}
	\hspace{-0.15in}
	{
		\includegraphics[width=0.24\textwidth]{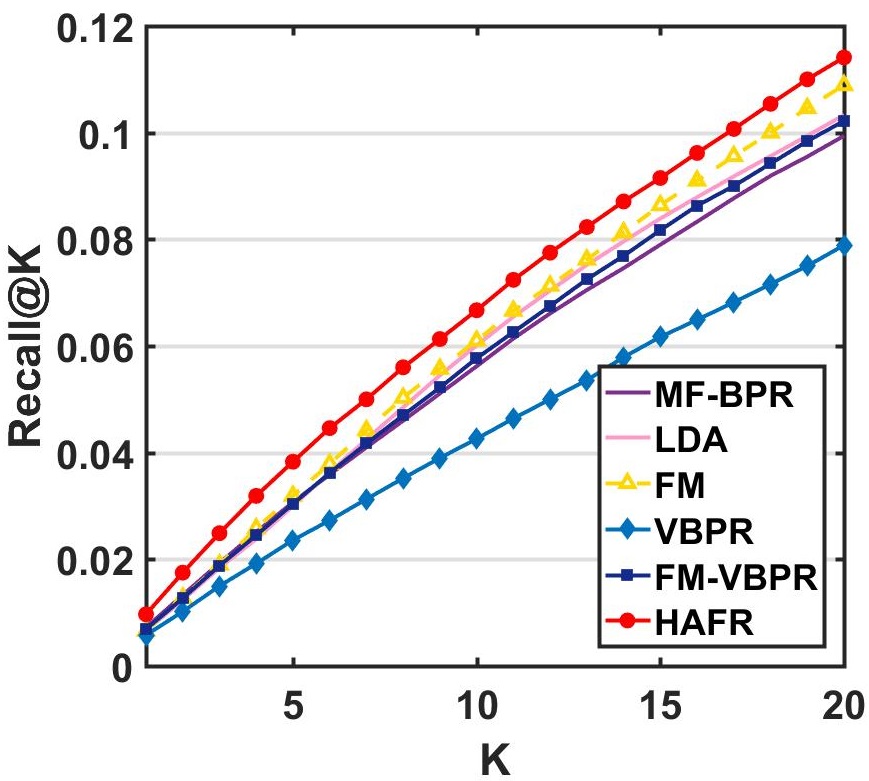}
		\label{fig:RQ_Recall}}
	\caption{Performance of Top-K recipe recommendation.}
	\label{Fig:Sparse}
	\vspace{-0.4cm}
\end{figure}

\subsection{Model Comparison (RQ1)}
Table \ref{tab:model performance} displays the performance comparisons across different models. Note that HAFR-non-v and HAFR-non-i are two variants of HAFR without input of recipe image and ingredients, respectively. From Table \ref{tab:model performance}, we have the following observations:
\begin{itemize}
    \item Our HAFR significantly outperforms other compared methods. This justifies the utility of HAFR that leverages the strong representation capability of multi-layer neural network and the hierarchical attention to jointly model user-recipe interactions and recipe contents (\ie image and ingredients).
    \item Specifically, compared to FM-VBPR - the state-of-the-art model, which has the same input features as HAFR, including user-recipe interaction, recipe image and recipe ingredients, HAFR exhibits an average improvement of 12\%. This is very remarkable, since FM-VBPR explores interactions between user, recipe, image and ingredients, which also integrate these key factors into the framework. 
    \item Compared to HAFR-non-v and HAFR-non-i, HAFR outperforms them by 24\% and 8\% on average, respectively. This illustrates that visual features and ingredient information are helpful for food recommendation because these side information can be informative and complement each other. This is also the reason why FM-VBPR, which additionally incorporates visual features, performs better than FM. 
    \item Among the baselines, FM-VBPR performs the best, which verifies the advantage of jointly modeling of historical interactions and recipe contents. VBPR and FM performs weaker since they lack ingredients and visual features, respectively. Lastly, MF and LDA, that model user-recipe interactions only, performs further worse, which justifies the benefit of recipe context once again. 
\end{itemize}
Note that NDCG and Recall metrics are not consistent with AUC metric. The reason may be that AUC is consistent with BPR loss, whereas NDCG and Recall focus more on recipe positions.

Figure \ref{Fig:Sparse} shows the performance of Top-K recipe recommendation where the ranking position K ranges from 1 to 20. As can be seen, HAFR consistently achieves the best performance on both NDCG and Recall metrics. 
This illustrates the robustness of HAFR and effectiveness of attention mechanism to explore information comprehensively. VBPR, as the only model considering both interactions and recipe image, performs unexpectedly bad, of which the reason is left for future exploration. 

\subsection{Effect of the Hierarchical Attention (RQ2)}
To better understand the proposed HAFR model, we further evaluate its key component---the \textit{Hierarchical Attention}, which consists of \textit{ingredient-level} and \textit{component-level} attention modules. Specifically, we compare two variants of HAFR by removing the ingredient-level attention (and the component-level one). It should be noted that the aggregation operation without the associated attention module (\eg aggregating ingredient embeddings into the representation of an ingredient list) will be deteriorated to a common average pooling operation. Table \ref{tab:model performance-attention} shows the performance of HAFR and its variants, from which, we have the following key observations:
\begin{itemize}
    \item 
    When the component-level attention is applied, the performance is improved (by 4\% on average) as compared with utilizing average pooling. The good performance signifies that users' preference over recipes comes from different components (\ie eating history, appearance, and ingredients), and justifies the effectiveness of dynamically aggregating the recipe information.
    \item By further applying the ingredient-level attention, relative performance improvement increases to 15\%. The further performance improvement demonstrates the benefit of dynamically awarding ingredients, which provide clues to infer users' preference over recipes. Besides, it indicates that an attention working at fine-grained (ingredient) level could benefit the attentive fusion at coarse-grained (component) level, which justifies the advantage of \textit{Hierarchical Attention}.
\end{itemize}

\begin{table}[t]
	\begin{center}
	\small
		\caption{Effect of attentions on ingredient (Ingre) and component (Comp) level. AVG and ATT represents the average pooling and attention, respectively. $^*$ denotes the statistical significance for $p < 0.05$.}\label{tab:model performance-attention}
		\renewcommand{\arraystretch}{1.05}
		\setlength{\tabcolsep}{1.4mm}{
		\begin{tabular}{|c|c|c|c|c|c|}
			\hline
			\bf{Model} &\multicolumn{2}{|c|}{\bf{Level}}   & \multicolumn{3}{|c|}{\bf Allrecipes}   \\\hline
			\multirow{4}{*}{HAFR} & Ingre & Comp & \bf AUC     & \bf NDCG@10 & \bf Recall@10 \\\cline{2-6}
			
			& AVG & AVG   &0.5927  &0.0373 & 0.0576 \\\cline{2-6}
			& AVG & ATT   &0.6215  &0.0399 & 0.0582 \\\cline{2-6}
			& ATT & ATT   &$\bf{0.6435}^*$ &$\bf{0.0455}^*$ &$\bf{0.0674}^*$\\\hline
		\end{tabular}}
		\vspace{-0.4cm}
	\end{center}
\end{table}
\begin{figure}[t]
	\centering
	\includegraphics[width=0.4\textwidth]{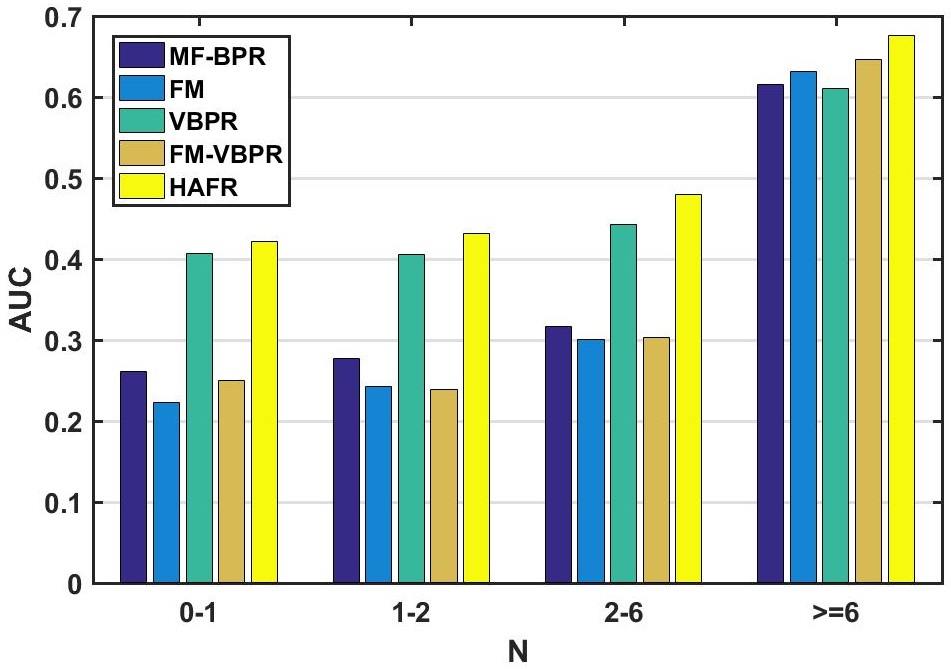}
	\caption{Performance over recipes of different popularity.}
	\label{fig:item_pop_performance}
	\vspace{-0.4cm}
\end{figure}
\subsection{Performance over Recipes of Different Popularity Levels (RQ3)}
Recall that the recipe popularity follows a long tail distribution. We then investigate the performance of models over recipes of different popularity levels, \ie we separately evaluate a trained model on recipes with different number of historical ratings. Specifically, we divide the test set into four groups of equal size based on the number of ratings per recipe ($N$). Figure \ref{fig:item_pop_performance} illustrates the performance of the compared models, from which we observe:
\begin{itemize}
    \item Our model HAFR with \textit{hierarchical attention} consistently outperforms other baseline methods in all the groups. The improvement demonstrates the robustness and capacity of HAFR, which may be because HAFR learns better food representation with the hierarchical attention, which thoroughly explores the interaction and context information of the recipe.
    \item We can see a clear trend that the performance of the models increases with more ratings per recipe. The reason may be that for popular recipes, there are more user-recipe historical information. Therefore, the models could capture user preferences more easily.
\end{itemize}

\begin{figure*}[t]
	\centering
	\subfloat[AUC]{
		\includegraphics[width=0.3\textwidth]{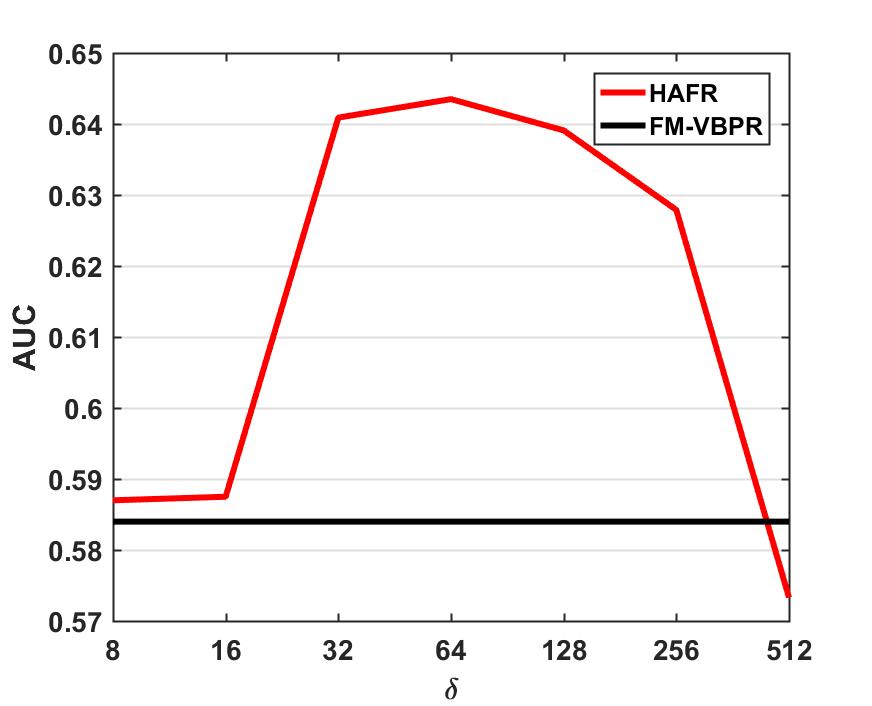}
		\label{fig:att-size_AUC}}
	\hspace{-0.15in}
	\subfloat[NDCG@10]{
		\includegraphics[width=0.3\textwidth]{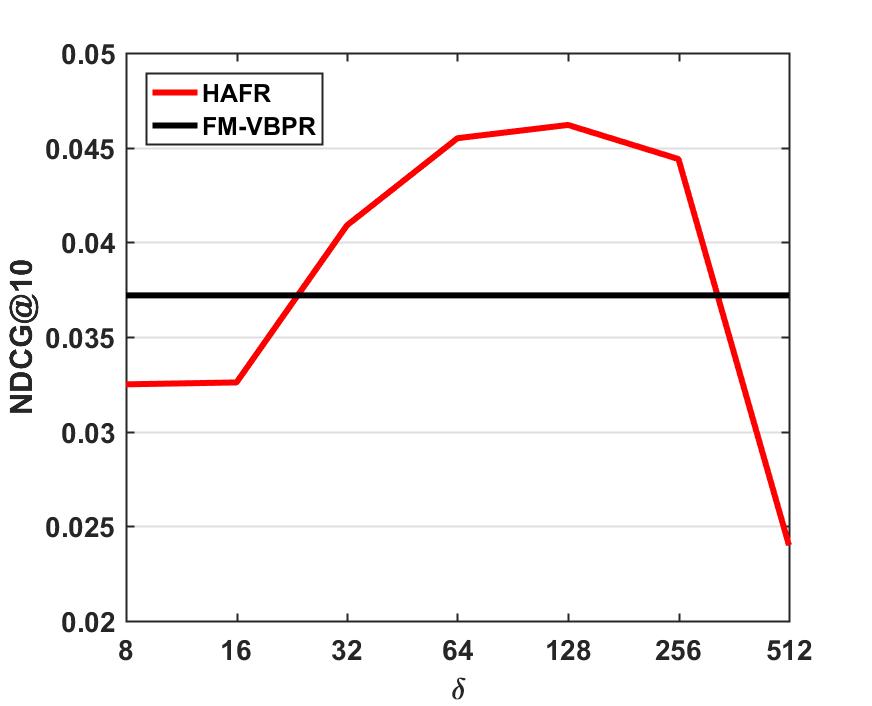}
		\label{fig:att-size_NDCG}}
	\hspace{-0.15in}
	\subfloat[Recall@10]{
		\includegraphics[width=0.3\textwidth]{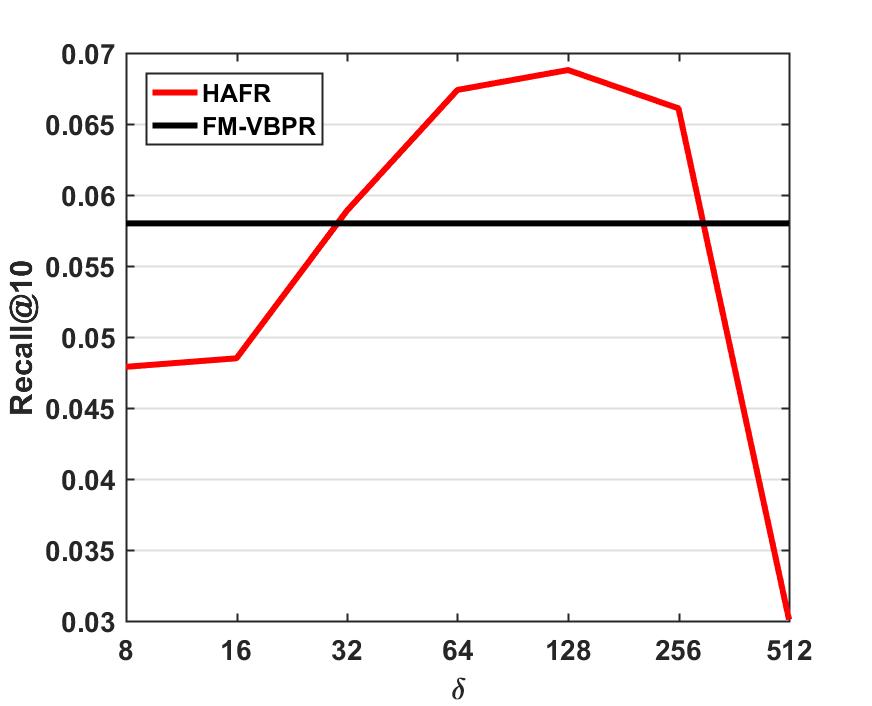}
		\label{fig:att-size_Recall}}
	\vspace{-0.2cm}
	\caption{Performance of HAFR regarding AUC, NDCG@10, and Recall@10 as adjusting the size of hierarchical attention. Larger value of $\delta$ means larger attention models with more parameters.}
	\label{fig:att_size_performance}
	\vspace{-0.4cm}
\end{figure*}

\begin{figure*}[t]
	\centering
	\subfloat[AUC]{
		\includegraphics[width=0.3\textwidth]{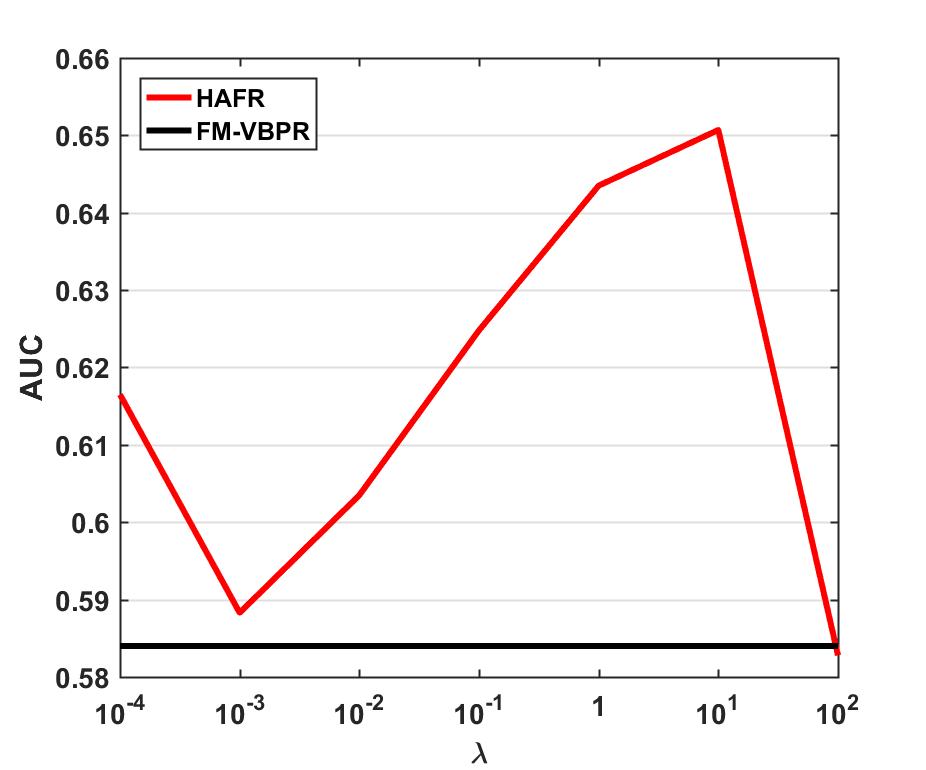}
		\label{fig:regh_AUC}}
	\hspace{-0.15in}
	\subfloat[NDCG@10]{
		\includegraphics[width=0.3\textwidth]{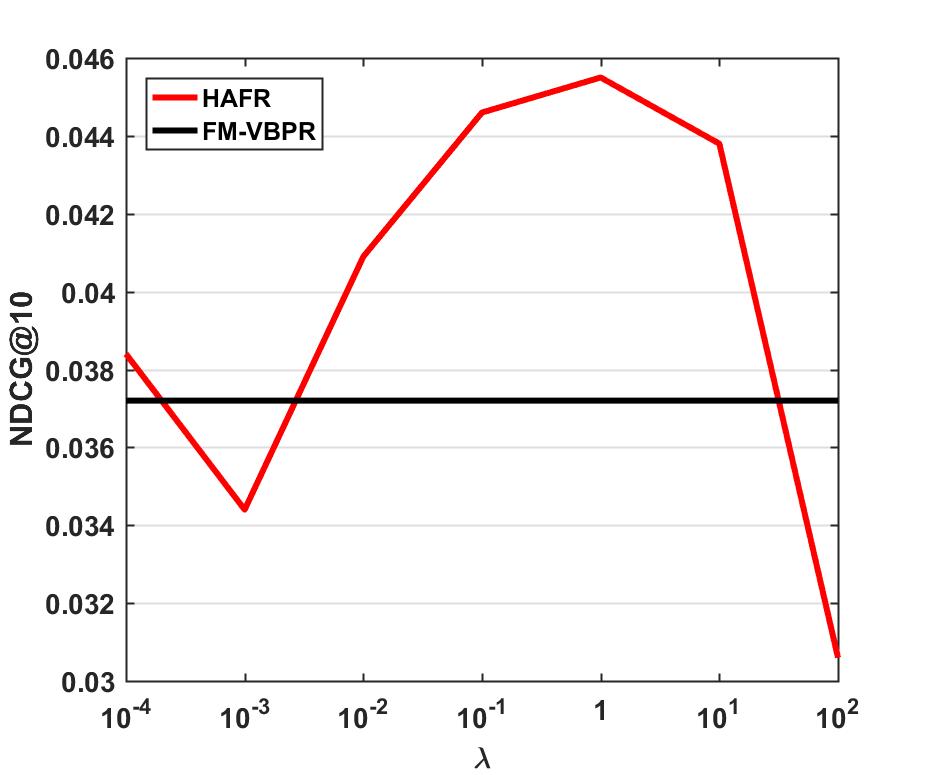}
		\label{fig:regh_NDCG}}
	\hspace{-0.15in}
	\subfloat[Recall@10]{
		\includegraphics[width=0.3\textwidth]{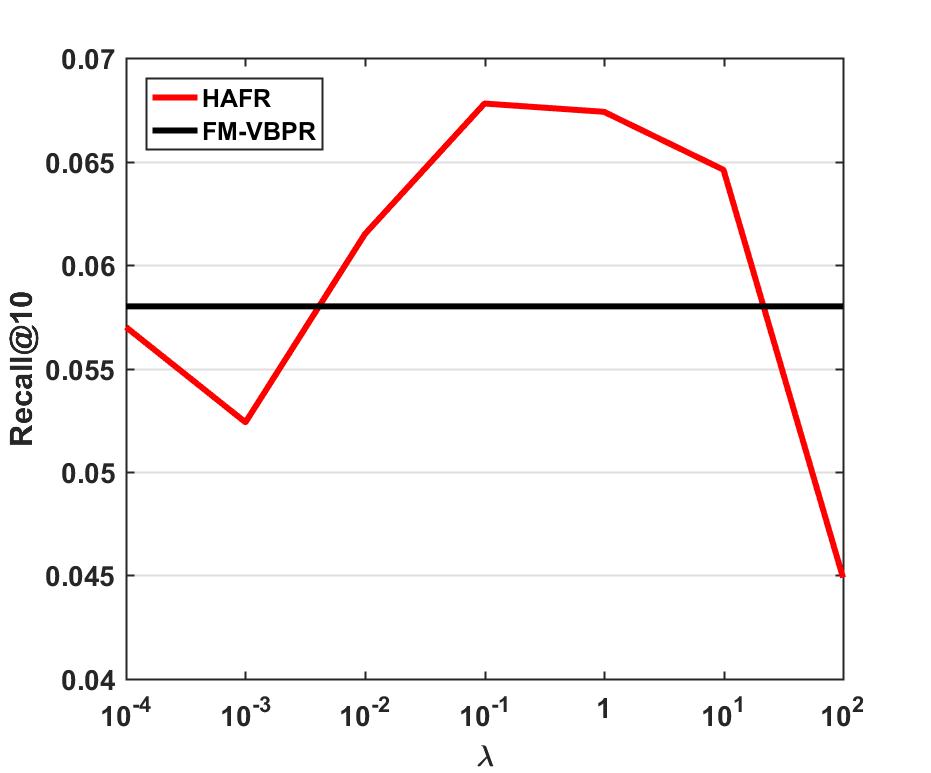}
		\label{fig:regh_Recall}}
	\vspace{-0.2cm}
	\caption{Performance of HAFR regarding AUC, NDCG@10, and Recall@10 as varying the strength of regularization applied on the parameters of output layer. Larger value of $\lambda$ indicates stronger regularization.}
	\label{fig:reg_h_performance}
	\vspace{-0.1cm}
\end{figure*}
\subsection{Hyper-parameter Study (RQ4)}
Hyper-parameters have significant impact on the performance recommender system~\cite{he2017neural,rendle2012factorization,liu2018discrete}. Therefore, we study the effect of hyper-parameters on the performance of HAFR. Specifically, since the introduced hierarchical attention is the core component of HAFR, the complexity of the hierarchical attention plays a pivotal role in affecting the performance of HAFR. To explore the impact, we investigate the model performance with different complexity by adjusting the size of hidden layer (\ie the number of hidden units) in the ingredient-level and component-level attentions. Note that we set the two attentions with the same number of hidden units to simplify the procedure. In addition, we study the influences of regularization terms on our HAFR method by taking the output layer as an example. That is, we study the model performance as we change the value of $\lambda$, which adjusts the strength of regularization ($L_2$-norm) applied on the parameters in the output layer. It should be noted that the output layer is selected because of its direct impact on the predictions.


\subsubsection{Complexity of Hierarchical Attention}
Figure \ref{fig:att_size_performance} shows the performance of HAFR with respect to different number of hidden units in the hierarchical attention (named as $\delta$), of which higher value means more model parameters and higher model complexity. Note that we depict the performance of FM-VBPR for reference. Clearly, regarding the three evaluation metrics, we can see a consistent trend that the performance of HAFR increases with more hidden units, \ie larger attention models. The results indicate that a larger model is beneficial to recommendation due to the increased modeling capability. However, the performance decreases when $\delta$ is larger than 128, which indicates that oversized attention may cause overfitting and degrade the performance. These findings are consistent with previous works such as~\cite{he2017neural,wang2018tem}. Moreover, HAFR consistently outperforms FM-VBPR with significant improvements when the value of $\delta$ is close to the optimal setting. It further demonstrates the effectiveness of the proposed hierarchical attention.
\subsubsection{Strength of Regularization} Figure \ref{fig:reg_h_performance} presents the performance of HAFR \wrt AUC, NDCG@10, and Recall@10 as we adjust the value of $\lambda$, which balances the training loss and regularization applied on the parameters of the output layer. As can be seen, when $\lambda$ is smaller than $10^{-2}$, HAFR's performance is much worse than the optimal setting. This indicates that HAFR could suffer from overfitting, which has also been shown in previous neural network-based recommender systems~\cite{he2017neural}. However, when the regularization term exceeds the optimal setting, the performance decreases, which is also reasonable since too strong regularization would prevent the model to learn from the training data. Moreover, when the value of $\lambda$ is extremely large ($10^2$), the performance of HAFR is worse than that of FM-VBPR with the optimal setting. Noting that the representation ability of FM-VBPR is weaker than HAFR, the results further demonstrate the importance of setting suitable regularization on neural network-based recommender systems.

%% file: 5.relatedwork.tex
\section{Related Work}\label{sec:related}
In this section, we discuss \textit{food recommendation} methods that are related to the work, followed by a summarization of studies on \textit{visually-aware recommendation}. In addition, since food recommendation is a specific application of food content analysis, we further discuss the recent work on \textit{food content analysis}. 

\subsection{Food Recommendation} 
Food recommendation has received a lot of attentions in recent years. Technically speaking, existing work can be divided into two categories: \textit{collaborative filtering} and \textit{content-based filtering} approaches. Collaborative filtering approaches focus on mining users' preference over recipes purely from user-recipe interactions with general CF algorithms. \cite{ge2015using} presents a matrix factorization approach for food recommendation which integrates user tags and ratings information to obtain better prediction accuracy than standard matrix factorization baselines. \cite{trattner2017investigating} tests various collaborative filtering methods for food recommendation, where the LDA approach performs best among all approaches. In LDA, users are regarded as documents and recipes as words. However, CF-based methods ignore the rich content of recipes, which is of vital importance for food recommendation.

Distinct from collaborative filtering approaches, content-based filtering methods only explore 
food contents for making recommendation. Regarding the types of considered contents, researches in this line can be further subdivided into \textit{ingredient-based}, \textit{image-based}, and \textit{profile-based} ones.
1) Ingredient-based methods make recommendations by mining user preference from recipe ingredients~\cite{freyne2010intelligent,freyne2010recommending,teng2012recipe}. 
For instance, \cite{teng2012recipe} take ingredients as recipe features, which are fed into a tree-based classifier to predict rating on a recipe from the target user.
2) Image-based methods show that algorithms designed to extrapolate important visual aspects of recipe images would enhance food recommendation~\cite{yang2015plateclick,yang2017yum}. 
3) Besides recipe contents, some studies show that side information of user, including gender~\cite{rokicki2016plate}, cultural background~\cite{kim2016tell} and other demographic factors~\cite{rokicki2017editorial}, also reflects users' characteristics over food choices. However, none of the existing methods jointly model the user-recipe interactions and key contents of recipe.

Since suitable food plays a vital role in avoiding health problems and improving nutritional habits, there have also been efforts explored healthy factors for food recommendation. \cite{ge2015health} incorporates nutritional aspects into the recommendation approach directly by accounting for calorie counts. \cite{elsweiler2015bringing} tries to obtain trade-off between what users want and nutritionally appropriate intake. The method combines these two aspects linearly as framework for food recommendation task.  
\cite{elsweiler2017exploiting} proposes a post-filtering approach to incorporate nutritional aspects in nutrition metrics which are based on widely accepted nutritional standards from The World Health Organization (WHO) \cite{amine2003diet}. 
We leave this extension as future work, as this work focuses on integrating the comprehensive food contents for food recommendation. 

\subsection{Visually-aware Recommendation}
With the development of image processing techniques, great efforts from both the industry and academia have been paid on visually-aware recommendation, in which the rich visual features of the items (\eg clothes, movies, images and videos posted in social media) are encoded to enhance the representation learning of users and items and improve the performance of recommender system.

In visually-aware recommendation, visual content associated with the item is mainly viewed as side information of the item to be recommended. Specifically, visual features are extracted from the visual content via deep neural networks, such as ResNet. The visual features are then incorporated into the conventional collaborative filtering framework as side information of the associated item. For instance, Wang \textit{et al.} extracted visual features from the image of point-of-interests (POIs) with VGG16, which is incorporated into the Matrix Factorization solution for POI recommendation~\cite{wang2017your}.
Yu \textit{et al.} extracted visual features from the image of clothes with CNN, then utilized Brain-inspired Deep Network to obtain aesthetic features, both of which are interacted into Collaborative Filtering model for clothing recommendation~\cite{DBLP:www/YuZHC0Q18}.

In addition to enhance the representation of targeted item, visual content of the items a user has interacted with is also used to enrich the representation of user for better inferring of user preference~\cite{chen2017attentive,DBLP:mm/Gelli0CC17}. For instance, 
Chen \textit{et al.} used hierarchical attention networks to learn user representation from the images the user has interacted with. HAN first aggregates the regional features of an image and then fuses the feature of images a user has interacted with to represent the user~\cite{chen2017attentive}.


Our work also aims to enhance the representation of item with visual features extracted from recipe image. However, different from the targeted items in previous work, \eg POIs and clothes, food also has key content information of its ingredients, which is closely related to the appearance of food. As such, we devise a hierarchical attention to aggregates these contents, which makes HAFR different from previous solutions.
\subsection{Food Content Analysis}
A lot of works focus on food-oriented applications via analyzing the content of recipes such as \textit{food categorization} and \textit{food retrieval}.

Food categorization aims to classify food into the correct category from its image input~\cite{herranz2017modeling, min2018you}. Technically speaking, food categorization approaches can be roughly divided into two main directions: \textit{single-label food categorization} and \textit{multi-label food categorization}, regarding the number of dishes in the input image. Single-label food categorization methods assume that only one dish is presented in the image and solve the task as a single-label classification~\cite{martinel2015structured}. On the contrast, multi-label food recognition focuses on images containing multiple dishes, which is solved as a multi-label classification~\cite{aguilar2018grab}. As formulated as classification problems, existing works under the two directions focus on extracting features from food images. Traditional solutions rely on hand-crafted features, such as SIFT, Garbor and LBP~\cite{martinel2015structured}. Recently, inspired by the success of deep learning techniques, deep neural networks, especially CNNs are exploited to learn food representations~\cite{kagaya2014food, martinel2018wide}. 

According to retrieval targets, existing work in food-related retrieval can be divided into: \textit{food image retrieval} and \textit{cross-modal recipe retrieval} methods. 

The target of food image retrieval is to find food images matched with the query could be either an image or a textual description.  Kitamura \textit{et al.} developed the first food image retrieval system, FoodLog, which supports image-based query only~\cite{kitamura2009foodlog}. 
Later on, Farinella \textit{et al.} combined different types of features, such as SIFT and Bag of Textons, to represent food images to further improve the retrieval performance~\cite{farinella2016retrieval}. Recently, Ciocca \textit{et al.} applied CNN-based features, which achieves state-of-the-art performance~\cite{ciocca2017learning}. 


Apart from image, a recipe could also contains textual descriptions including its ingredients and cooking instructions. Cross-model recipe retrieval is to retrieve the textual description of recipes from image-based query~\cite{chen2016deep,chen2017cross,min2017being,chen2018deep}. Among the existing solutions, the ones focus on representation learning of the query and recipe achieve promising performance, where the matching between query and recipe are modeled on the learned representations. For instance, Chen \textit{et al.} encoded the image and text into the same multi-dimensional space via CNN and RNN equipped with attention modeling~\cite{chen2018deep}. 